\documentclass[aps,prd,twocolumn,a4paper]{revtex4}
\usepackage{ulem}
\usepackage{color}
\usepackage{graphicx}
\usepackage{epsfig}
\newcommand{\be}{\begin{equation}}
\newcommand{\ee}{\end{equation}}
\newcommand{\ba}{\begin{eqnarray}}
\newcommand{\ea}{\end{eqnarray}}

\begin{document}
\title{Heavy flavor suppression: role of hadronic matter} 
\author{Santosh K. Das$^{a,b}$, Sabyasachi Ghosh$^c$, Sourav Sarkar$^a$ and Jan-e Alam$^a$}

\affiliation{$^a$ Variable Energy Cyclotron Centre, 1/AF, Bidhan Nagar , Kolkata - 700064}
\affiliation{$^b$ Department of Physics and Astronomy, University of Catania, 
Via S. Sofia 64, 1-95125 Catania, Italy}
\affiliation{$^c$ Centre for Astroparticle Physics and Space Bose Institute Block-EN, Sector-V, Salt Lake
Bidhan Nagar, Kolkata - 700091 India}

\begin{abstract}
The role of hadronic matter in the suppression of open heavy flavored 
mesons has been studied. The heavy-quarks (HQs) suppression factors have been calculated 
and contrasted with the experimental data obtained from nuclear collisions at Relativistic Heavy Ion Collider
(RHIC) and Large Hadronic Collider (LHC) experiments. It is found that the 
suppression in the hadronic phase at RHIC energy is around $20\%-25\%$ whereas 
at the LHC it is around $10\%-12\%$ 
for the D meson. In case of B meson the hadronic suppression is around $10\%-12\%$ 
and $5\%-6\%$ at RHIC and LHC energies respectively. Present study suggests that the 
suppression of heavy flavor in the hadronic phase is  significant 
at RHIC. However, the effect of hadronic suppression at LHC is marginal, this
makes the characterization of QGP at LHC less complicated. 
  
\vspace{2mm}

\end{abstract}
\maketitle

One of the  primary aims of the ongoing heavy-ion collision experiments at 
RHIC and LHC energies is to create and study the properties of Quark Gluon Plasma (QGP).
The heavy flavours (HF) play a vital role to serve this purpose
~\cite{mooreteaney,rapp, ko, adil, gox,hatsuda,das,mazumder,das1, bib, alberico}.
In particular the depletion of high transverse momentum $(p_T)$ hadrons (D and B) produced 
in Nucleus + Nucleus collisions relative to those produced in proton
+ proton (p+p) collisions has been considered as an indicator of QGP formation. 
The STAR~\cite{stare}, PHENIX~\cite{phenixe} and the ALICE~\cite{alice} 
collaborations have measured this high $p_T$ depletion.
To make the characterization of the QGP 
reliable the role of the hadronic matter should be taken into 
consideration and its contribution must be disentangled  from the 
experimental observables. In this work an attempt has been made to estimate the 
effect of the hadronic phase on the nuclear suppressions of HFs.

We study the evolution of the HFs in the following scenario.
We assume that the light quarks, anti-quarks and gluons form a thermalized 
matter and the non-equilibrated heavy quarks(HQs) 
are moving through the expanding QGP background. 
While the evolution of the expanding QGP is described by the
relativistic hydrodynamics with initial temperature 
and the thermalization time constrained by the measured
charged particle multiplicity, 
the motion of the non-equilibrated HQ is described by the 
Fokker-Planck equation (FP) with drag and diffusion co-efficients 
arising due to the interaction of HQs with the expanding QGP background. 
The initial conditions for the distributions of HQs have been 
taken from the  NLO pQCD results obtained for pp collisions 
by using the MNR code~\cite{mnr}. 

The expanding QGP converts to hadronic system when it cools down to the
transition temperature, $T_c$. The solution of the FP equation
for the charm and bottom quarks  
at the transition point is folded with the Peterson fragmentation function 
~\cite{peterson} to obtain the momentum distributions of the heavy 
flavoured mesons containing the effects of the interaction of the expanding QGP background.
The hadronic matter evolves in space and time described  
by relativistic hydrodynamics till the matter gets dilute enough
to freeze-out kinematically. 
The motion of the non-equilibrated HF mesons ($D$ and $B$) in the
expanding hadronic system is again described by the FP equations with
drag and diffusion coefficients  evaluated due to their
interactions with  hadronic matter.
The solution of the FP equation for the $D$ and $B$ 
mesons at the freeze-out point encompassing the effects of drag of both 
the QGP and the hadronic phases has been used to determine the 
suppression in the high $p_T$ domain.

The FP equation describing the motion of the non-equilibrated degrees of freedom (dof)
in the bath of the equilibrated dof reads as ~\cite{sc,svetitsky},  
\be
\frac{\partial f}{\partial t} = 
\frac{\partial}{\partial p_i} \left[ A_i(p)f + 
\frac{\partial}{\partial p_j} \lbrack B_{ij}(p) f\rbrack \right] 
\label{FP}
\label{eq1}
\ee
where $f$ is the momentum distribution of the 
non-equilibrated dof, $A_i$ and $B_{ij}$ are related to the drag and diffusion coefficients. 
The interaction between the probe and the medium enter through the 
drag and diffusion coefficients. 

During their propagation through the QGP the HQs dissipate
energy predominantly by two processes~\cite{das,mazumder,dks}: (i) collisional, {\it e.g.}  
$gQ \rightarrow gQ$,  $qQ \rightarrow qQ$ and $\bar{q}Q \rightarrow \bar{q}Q$
and (ii) radiative processes,  {\it i.e.} $Q+q\rightarrow Q+q+g$ and 
$Q+g\rightarrow Q+g+g$ . The  dead cone and Landau-Pomeranchuk-Migdal (LPM) effects on 
radiative energy loss of heavy quarks have been considered. 
Both radiative and collisional processes of energy loss are 
included in the effective drag and diffusion coefficients~\cite{das,mazumder}. 
The solutions of the FP equation have been convoluted with the 
Peterson fragmentation functions to obtain the D and B meson spectra at $T_c\sim 170$ MeV.
We omit the detailed description here as it is available in ~\cite{das,mazumder}.

The motion of the out-of-equilibrium HF mesons ($D$ and $B$) in the
expanding hadronic matter (HM) 
is studied by using the FP equations. 
The drag and diffusion coefficients 
of the $D$ and $B$ mesons have been calculated in~\cite{Ghosh,Das3}
for their interactions with pions, kaons, nucleons and eta)
(see also~\cite{laine,MinHe,abreu}) .
The $p_T$ distribution of the HF mesons obtained by convoluting the 
Peterson fragmentation function with the solution of the FP equation at the end of the QGP phase
has been used as  input (initial ) for solving the FP equation in the hadronic phase. 
The solution of the FP equation for the $D$ and $B$
mesons in the expanding HM at the freeze-out is employed to determine the
nuclear suppression. 
The expansion of the background medium (either QGP or HM) is
described by relativistic hydrodynamics~\cite{bjorken} with equation of state (eos)
which leads to the velocity of sound, $c_s=1/\sqrt{4}$.  

The suppression of high $p_T$ $D$ or $B$ mesons in the QGP phase, $R^Q_{AA}$,   
is given by:
$R^{Q}_{AA}=\frac{f_{Q}}{f_{i}}$
where $f_{Q}$ is given by the convolution of the solution of the FP equation 
at  the end of the QGP phase with the HQ fragmentation to $D$ or $B$ meson 
and $f_{i}$ is 
the function obtained from the convolution of the initial heavy quark momentum distribution 
with the HQ fragmentation function to $D$ and $B$  mesons. 
Similarly the suppression factor in the hadronic
phase alone can be written as,
$R^{H}_{AA}=\frac{f_{H}}{f_{Q}}$
$f_{H}$ is the solution of the FP equation describing the evolution in the
hadronic phase at the freeze-out. 
The net suppression of the HFs during the entire evolution process - from the 
beginning of the QGP phase to the end of the hadronic phase is given by:
$R_{AA}=R^{Q}_{AA}\times R^{H}_{AA}$

\begin{figure}[ht]
\begin{center}
\includegraphics[scale=0.40, clip=true]{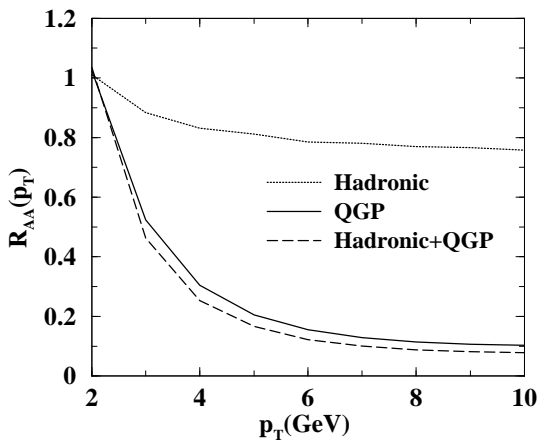}
\caption{Variation of $D$ meson suppression at RHIC energy for 
QGP, hadronic and hadronic+QGP phase. }
\label{fig1}
\end{center}
\end{figure}

\begin{figure}[ht]
\begin{center}
\includegraphics[scale=0.40, clip=true]{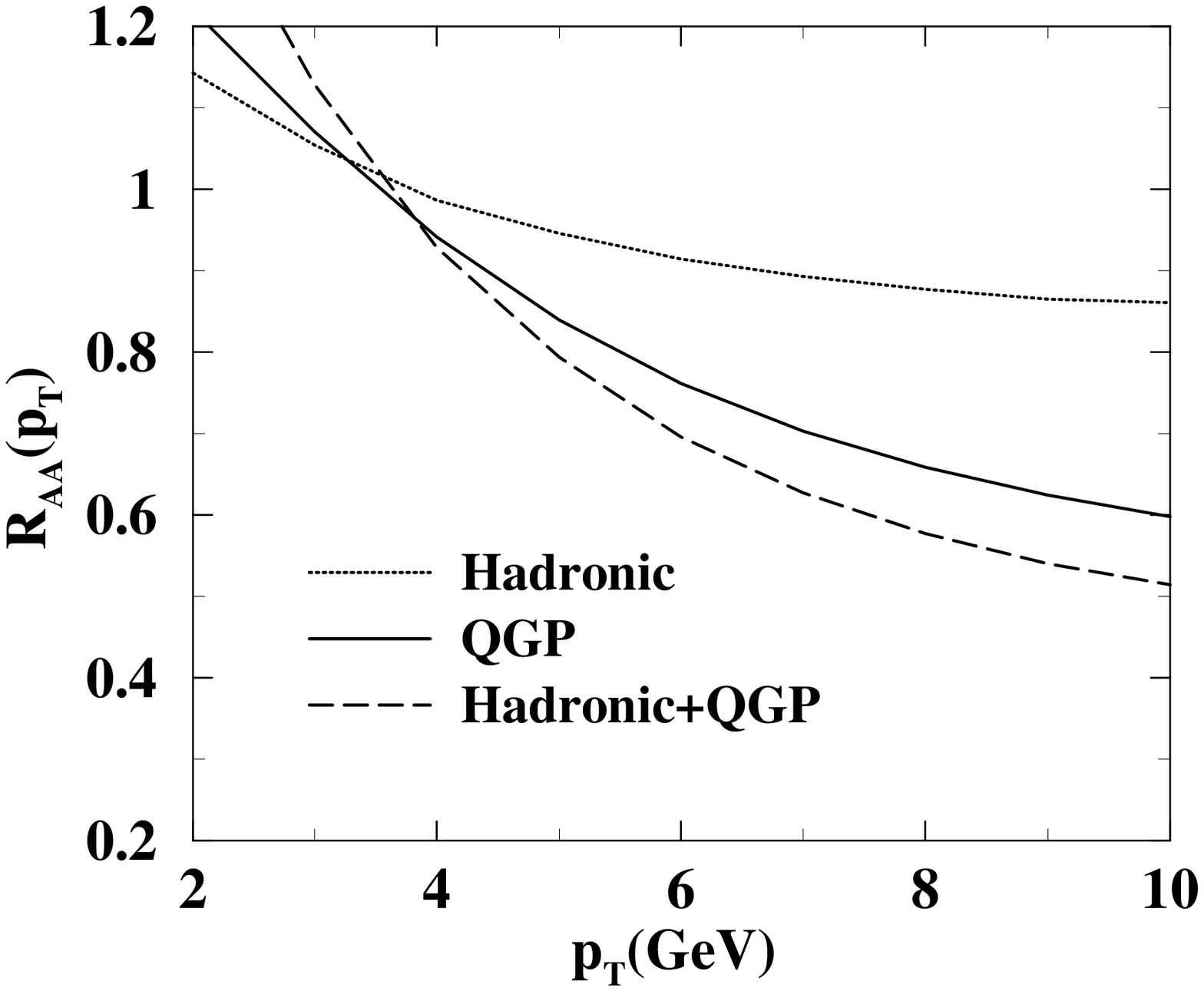}
\caption{Variation of $B$ meson suppression at RHIC energy for 
QGP, hadronic and hadronic+QGP phase. }
\label{fig2}
\end{center}
\end{figure}

\begin{figure}[ht]
\begin{center}
\includegraphics[scale=0.30, clip=true]{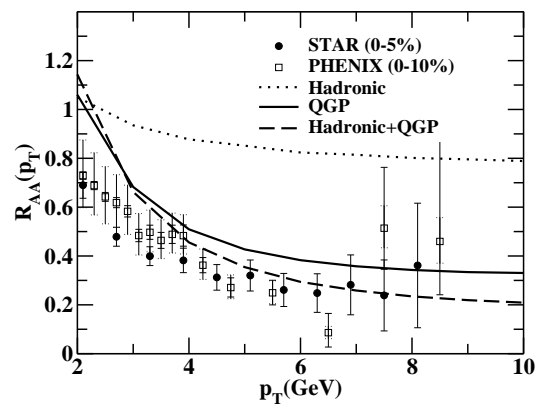}
\caption{ 
$R_{AA}$ of heavy flavoured mesons measured through their
semi-leptonic decays as a function of $p_T$ at RHIC (see text).
Experimental data taken from ~\cite{stare} and ~\cite{phenixe}.
}
\label{fig3}
\end{center}
\end{figure}
 
\begin{figure}[ht]
\begin{center}
\includegraphics[scale=0.30, clip=true]{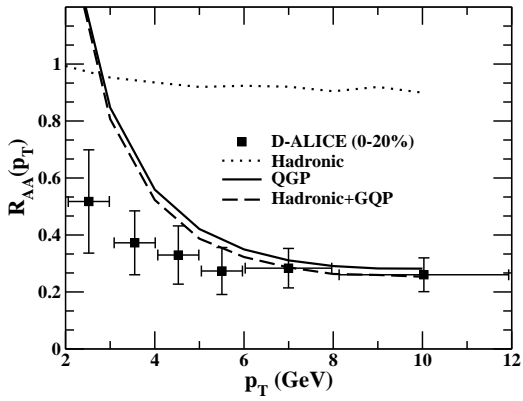}
\caption{
$R_{AA}$ of $D$ meson as a function of  $p_T$ at LHC.
Experimental data taken from ~\cite{alice}.
}
\label{fig4}
\end{center}
\end{figure}


\begin{figure}[ht]
\begin{center}
\includegraphics[scale=0.40, clip=true]{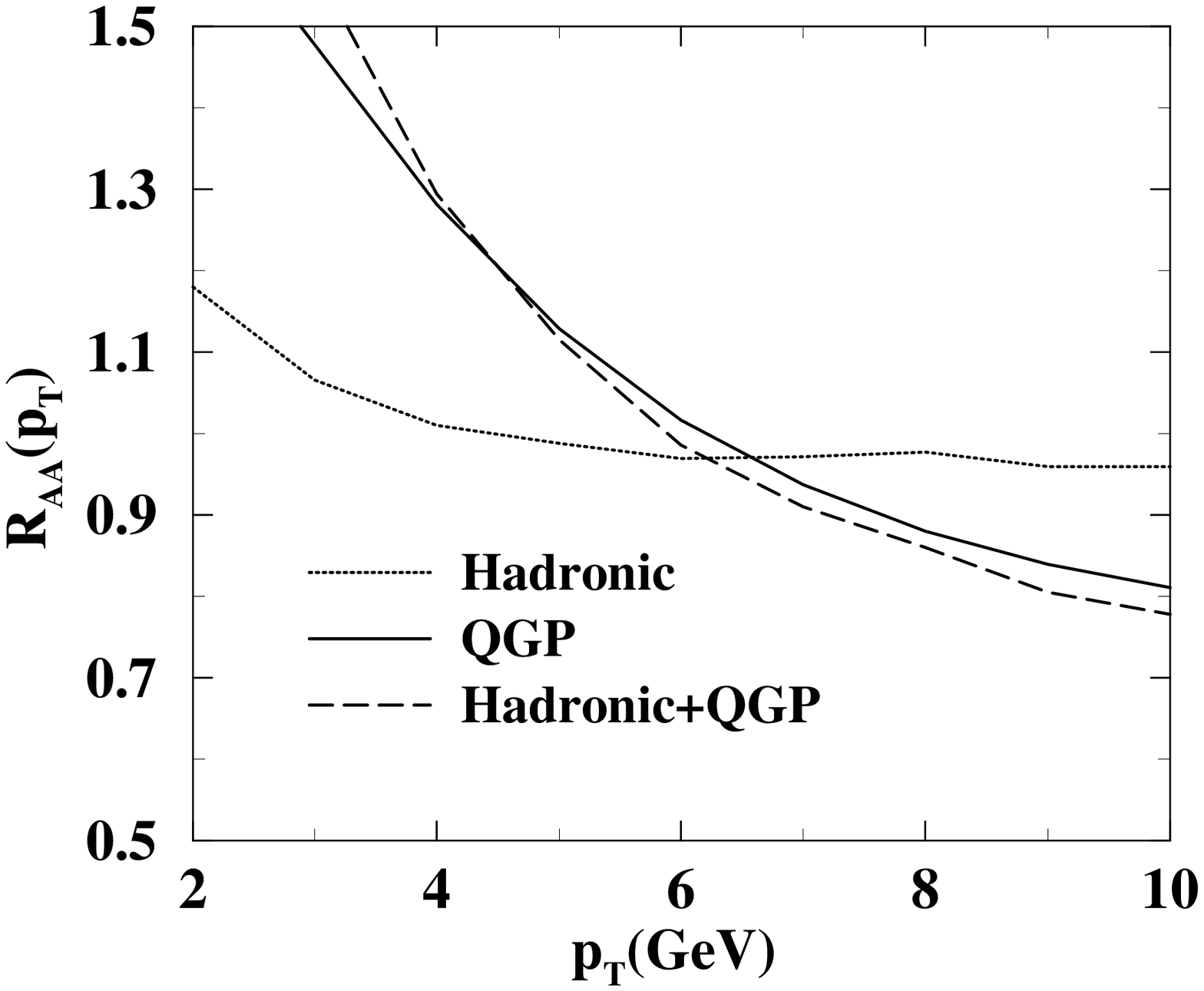}
\caption{Variation of $B$ meson suppression at LHC energy for 
QGP, hadronic and hadronic+QGP phase. }
\label{fig5}
\end{center}
\end{figure}

\begin{figure}[ht]
\begin{center}
\includegraphics[scale=0.30, clip=true]{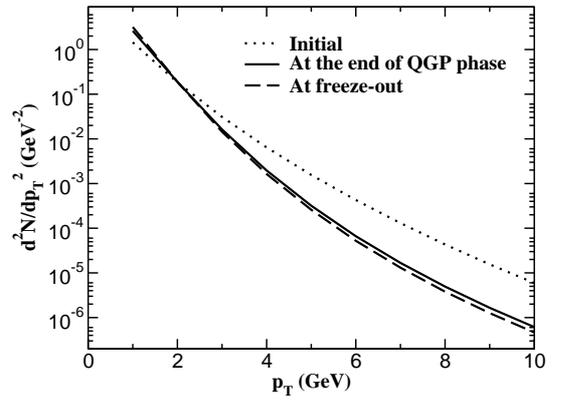}
\caption{The  $p_T$ spectra of the $D$ mesons at RHIC obtained
by convoluting the charm quark to $D$ meson fragmentation with (i) the initial charm quark distribiuon 
(dotted line), (ii) solution of the FP equation at the transition point (solid line) and (iii) the solution of the FP equation
at the end of the hadronic phase (i.e. at freeze-out) which 
contains the effects of suppression in the QGP as well as hadronic phases (dashed line). 
}
\label{fig6}
\end{center}
\end{figure}

\begin{figure}[ht]
\begin{center}
\includegraphics[scale=0.30, clip=true]{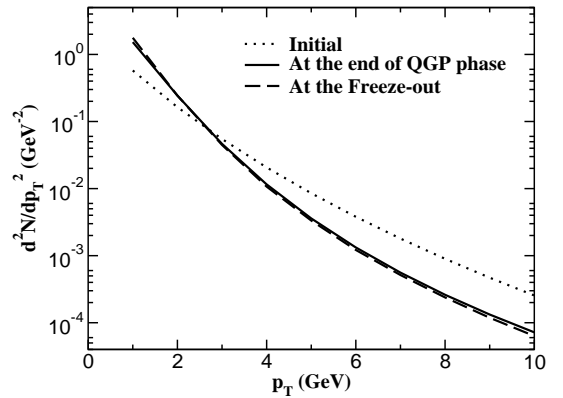}
\caption{Same as Fig.~\ref{fig5} for LHC conditions.
}
\label{fig7}
\end{center}
\end{figure}

The results for the $D$ meson at RHIC energy is depicted in Fig.~\ref{fig1}. 
We have taken the initial 
temperature, $T_i=0.4$ GeV and thermalization time, $\tau_i=0.2$ fm/c.
These values are constrained by the measured hadronic multiplicity, $dN/dy=1100$.
We observe that the D meson suppression 
in the hadronic phase is around $20\%-25\%$ for $p_T=3-10$ GeV at RHIC energy. 
This suggests that the effects of the hadronic medium  
on the charmed meson suppression is non-negligible. 
Therefore, these effects should be excluded from the experimental
data to estimate the suppression in QGP and make the characterization of QGP
definitive.  
The results for $B$ meson is displayed in Fig~\ref{fig2} at RHIC. In the hadronic phase the 
$B$ meson suppression is around $10\%-12\%$, indicating greater suppression of $D$ than $B$.  
However, the overall suppression of $B$ is also less than $D$. Because the drag of $b$ quarks
($B$ mesons) in QGP (HM) is smaller than that for $c$ quarks ($D$ meson).  

In Figs.~\ref{fig1} and ~\ref{fig2} the $R_{AA}$ have been plotted for 
$D$ and $B$ mesons individually for RHIC collision conditions. 
However, the data for $D$ and $B$ mesons are not available separately from  RHIC experiments. 
The PHENIX and STAR collaborations~\cite{phenixe,stare} have measured the $R_{AA}(p_T)$ of 
non-photonic single electrons originating from the decays of mesons containing 
both open charm and bottom quarks, {\it i.e.} the experimental data contains
suppression of both the charm and bottom through the measured, $R_{AA}(p_T)$.
Theoretically $p_T$ spectra of non-photonic electrons originating from the decays of $D$ and
$B$ mesons ($D\rightarrow X e \nu$ and $B\rightarrow X e \nu$)  
produced in heavy ion collisions have been obtained by following the procedure discussed in~\cite{das}). 
The $p_T$ spectra of single electrons originating from the pp collisions is accomplished 
by using the HQs distributions obtained from the MNR code.
The ratio of electron spectra from heavy
ion to pp collisions gives $R^Q_{AA}$. Similar exercise has been performed  for the hadronic phase
to obtain $R^H_{AA}$. 

The theoretical results for QGP and hadronic phases along with the total
suppression is contrasted with the experimental data from RHIC experiments in Fig.~\ref{fig3}.
The results reveal that with the inclusion of the hadronic contributions the description
of experimental results improves. 
For LHC the experimental results on the $D$ suppression  is available directly~\cite{alice}. 
We have taken the value of $T_i=550$ MeV 
and $\tau_i=0.1$ fm/c  for $\sqrt{s_{NN}}=2.76$ TeV. 
It is found that the  $D$ meson suppression 
in the hadronic phase at LHC energy is around $10\%-12\%$ for $p_T=3-10$ GeV. 
The comparison of theoretical and experimental results (Fig.~\ref{fig4}) indicate that the hadronic phase 
play less dominant role at LHC than RHIC.  
It will be interesting to compare the experimental data on $B$
with the theoretical results and check whether both $D$ and $B$ spectra are reproduced simultaneously 
with same initial condition. 

It is to be noted that the theoretical descriptions over estimate the experimental data for $p_T\leq 3$ GeV
at RHIC and for $p_T\leq 4$ GeV  at LHC. The spectra of $D$ and $B$ mesons are obtained 
here from the fragmentation of high energy charm and bottom quarks. Such mechanisms of hadronization
may not be valid for the low $p_T$ hadrons. The low $p_T$ hadrons 
may be produced from the coalescence of thermal partons ~\cite{fries}. The 
$D$ and $B$ meson spectra at lower $p_T$ may be reproduced by 
using coalescence model calculations~\cite{grecohq}.

Fig.~\ref{fig5} displays the depletion of $B$ mesons at LHC.
The effects of the hadronic phase 
is found to be negligibly small. 
Indicating the fact that the response of the hadronic medium is less pronounced  
at LHC than RHIC. Therefore, the role of hadronic medium in characterizing 
the QGP by using heavy flavours can be ignored making the task of QGP 
detection less complex at LHC.

The differences in the magnitude of $R^H_{AA}$ in RHIC and LHC can be understood 
from the corresponding results plotted in Fig.~\ref{fig6} and Fig.~\ref{fig7}.   
The temperature of the hadronic system for both RHIC and LHC 
varies from $T_c$ to $T_f$ (170 to 120 MeV)  and therefore,
the value of the drag coefficients remain same.
However, the input distribution
to the hadronic matter 
is harder at LHC than RHIC, resulting in less suppression at LHC.

\begin{figure}[ht]
\begin{center}
\includegraphics[scale=0.30, clip=true]{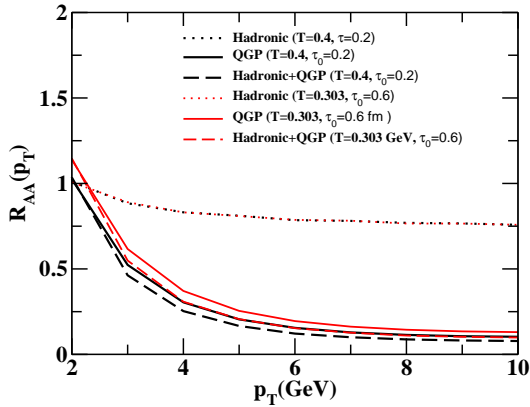}
\caption{(color on-line) The variation of $R_{AA}$ with $p_T$ 
for two sets of initial conditions.  
}
\label{fig8}
\end{center}
\end{figure}
Some comments on the sensitivity of $R_{AA}$ on the initial condition are
in order here.
The initial temperature and the thermalization time of the QGP is not uniquely known. 
Therefore, it may be interesting to study the sensitivity of $R_{AA}$ on the initial
conditions. In Fig.~\ref{fig8} we display the results for two sets of initial conditions
 keeping other parameters like $T_c$ and $T_f$ unaltered.
We observe that the suppression in the hadronic phase is negligible due to
the change in the initial conditions as expected because
the maximum ($T_c$ ) and the minimum ($T_f$ ) temperature of
this phase are kept unaltered.
However, the suppression in the QGP phase changes by approximately $20\%$ 
due to the change in the initial conditions as indicated in Fig.~\ref{fig8}. 
For higher $T_i$ the the drag in the QGP phase is higher which
results in more suppression. 
The net change (QGP+hadronic) also remains about $20\%$ as the change in
the hadronic phase can be ignored.

In summary we have evaluated the suppression  of HFs 
due to their interactions with the QGP and HM. 
While the HF suppression in QGP is used as a signal of QGP,
the hadronic suppression is treated as background. 
We observe that the suppression of $D$ is more than $B$ in the hadronic medium because
the hadrons drag the $D$ more than the $B$~\cite{Das3}.
The suppression at RHIC energy is significant and hence the hadronic contributions
should be taken into account in analyzing the experimental data. 
It is also interesting to note that the role of hadronic medium in HFs (especially for
$B$) suppressions at LHC is not substantial  
because $D$ and $B$ meson distributions are harder at LHC than RHIC.
Since the role of hadrons in $B$ meson suppression is very minimal, therefore, 
$B$ may play a unique role in characterizing QGP. This has a great advantage 
compared to other signals of QGP, for example, in case of electromagnetic 
probes (see~\cite{rw,alam1,alam2} for review) {\it i.e.} direct photons and
lepton pairs the role of hadronic matter is significant, which makes the  
task of extracting QGP properties difficult after filtering out hadronic contributions.  



\vspace{2mm}
SKD acknowledges the support by the ERC StG under the QGPDyn 
Grant No.259684

\end{document}